\documentclass[12pt]{article}
\usepackage{amsthm,amssymb,amsmath}
\usepackage{graphicx}
\newcommand{\hnn}{h\to\chi\chi}
\newcommand{\gev}{\mathrm{GeV}}
\newcommand{\tev}{\mathrm{TeV}}
\def\lsim{\raise0.3ex\hbox{$\;<$\kern-0.75em\raise-1.1ex\hbox{$\sim\;$}}}
\newcommand{\AddrLNF}{
  {\it INFN, Laboratori Nazionali di Frascati,C.P. 13, I00044 
    Frascati, Italy}}
\newcommand{\AddrUdeA}{
  {\it Instituto de F\'\i sica, Universidad de Antioquia,
    A.A. {\it{1226}}, Medell\'\i n, Colombia}}
\newcommand{\AddrUAM}{
  {\it Departamento de F\'{\i}sica Te\'orica C-XI, Universidad
    Aut\'onoma de Madrid, Cantoblanco, E-28049 Madrid, Spain
  }}
\newcommand{\AddrWU}{
  {\it Institut f\"ur Theoretische Physik und Astrophysik, 
    Universit\"at W\"urzburg, 97074 W\"urzburg, Germany 
  }}

\begin{document}
\begin{titlepage}
\begin{center}
\textbf{{\large Novel Higgs decay signals in\\
    \vspace{0.2cm}
    R-parity violating models}}\\[10mm]
D. Aristizabal Sierra$^a$,  W. Porod$^b$,  D. Restrepo$^c$, 
Carlos E. Yaguna$^d$
\vspace{0.4cm}\\
$^a$\AddrLNF.\vspace{0.4cm}\\
$^b$\AddrWU.\vspace{0.4cm}\\
$^c$\AddrUdeA.\vspace{0.4cm}\\
$^d$\AddrUAM.\vspace{0.4cm}
\end{center}
\begin{abstract}
  In supersymmetric models the lightest Higgs boson may decay with a
  sizable branching ratio into a pair of light neutralinos. We analyze
  such decays within the context of the minimal supersymmetric
  standard model with R-parity violation, where the neutralino itself
  is unstable and decays into Standard Model fermions. We show that
  the R-parity violating couplings induce novel Higgs decay signals
  that might facilitate the discovery of the Higgs boson at colliders.
  At the LHC, the Higgs may be observed, for instance, through its
  decay -via two neutralinos- into final states containing missing
  energy and isolated charged leptons such as $\ell^\pm\ell^\mp,
  \ell^\pm\ell^\pm, 3\ell$, and $4\ell$. Another promising possibility
  is the search for the displaced vertices associated with the
  neutralino decay. We also point out that Higgs searches at the LHC
  might additionally provide the first evidence of R-parity violation.
\end{abstract}
\end{titlepage}
\section{Introduction}
\label{sec:int}
The discovery of the Higgs boson is probably the most important goal
of the LHC. In the Standard Model, the Higgs branching ratios depend
only on the unknown Higgs mass, which is constrained to be larger than
$114~\gev$.  In the minimal supersymmetric Standard Model (MSSM) the
Higgs sector is more involved, as it includes two Higgs
doublets~\cite{Haber:1984rc}.  Yet, the mass of the lightest Higgs is
very restricted, typically below $135~\gev$. Another remarkable
feature of the MSSM Higgs is that it may decay into two light
neutralinos.  Besides being \emph{invisible}, such a decay causes a
suppression of all other branching ratios of the Higgs boson,
rendering more difficult its observation. In R-parity violating
models, however, a new twist occurs. There, the neutralino is unstable
and may decay within the detector into Standard Model fermions. Thus,
the decay $\hnn$, with $\chi$ subsequently decaying into light
particles, becomes \emph{visible} and new signals for Higgs decays at
colliders appear.  That is exactly the situation we aim to study in
this paper.

The most general supersymmetric version of the Standard Model is
phenomenologically inconsistent, for it includes lepton and baryon
number violating operators that would induce fast proton decay. In the
MSSM, an \emph{ad hoc} discrete symmetry known as R-parity is imposed
to prevent the decay of the proton. R-parity additionally implies the
conservation of lepton and baryon number as well as the stability of
the lightest supersymmetric particle --the LSP. Assuming R-parity,
however, is not the only way of preventing proton decay.  Lepton
parity and baryon triality~\cite{Ibanez:1991hv} are among the
alternative discrete symmetries that forbid proton decay but allow for
R-parity violation and, respectively, baryon or lepton number breaking
couplings. R-parity, therefore, is not an essential ingredient of low
energy supersymmetry.

Supersymmetric models with broken R-parity are well motivated
extensions of the Standard Model and have been amply considered in the
literature. They feature a rich phenomenology, markedly different from
that of the MSSM. The LSP, for instance, is unstable and consequently
it is no longer bound to be a neutralino; any supersymmetric particle
can be the LSP~\cite{Dimopoulos:1989hk}.  And since the LSP decays
into Standard Model fermions, distinctive decay patterns and collider
signals are expected
\cite{Porod:2000hv,Hirsch:2002ys,AristizabalSierra:2004cy}. R-parity
violation might also be at the origin of neutrino masses and mixing
~\cite{Hall:1983id}.  The bilinear R-parity violating model, in
particular, not only accounts for the observed values of neutrino
masses and mixing angles \cite{Hirsch:2000ef,Hirsch:2004he} but it
also predicts simple correlations between them and the LSP decay
branching
ratios~\cite{Mukhopadhyaya:1998xj,Choi:1999tq,Porod:2000hv,Hirsch:2002ys,AristizabalSierra:2004cy}.
These unique signals of R-parity violating models may soon be tested
at the LHC as well as at future colliders.

The mass of the lightest neutralino is not constrained by accelerator
searches or precision experiments \cite{Dreiner:2007fw,Chang:2008cw}.
If the GUT relation between gaugino masses is assumed, then the LEP
limit on the chargino mass, $m_{\chi^\pm}\gtrsim 100~\gev$
\cite{Yao:2006px}, translates into a lower bound on the neutralino
mass: $M_1\sim m_\chi\gtrsim 50~\gev$\cite{Yao:2006px}. If the
assumption of gaugino mass unification is not made, however, there is
no general limit on the mass of the lightest neutralino.  Thus,
neutralinos with masses below $50~\gev$, light neutralinos, are
perfectly compatible with present experiments.

For the MSSM, the implications of such light neutralinos in Higgs
boson decays were recently studied in \cite{Yaguna:2007vm}. There,
after examining the dependence of the $\hnn$ branching ratio with the
relevant supersymmetric parameters, it was shown that the decay into
two neutralinos can even be the dominant decay mode of the Higgs
boson, with a branching ratio as large as $80\%$. Here, we extend such
an analysis to supersymmetric models with R-parity violation. The
R-parity violating couplings may cause the decay of the neutralino
within the detector, transforming the former \emph{invisible} decay
$\hnn$ into a \emph{visible} one.

We will consider both bilinear and trilinear R-parity violating
operators and study the different 3-body neutralino decays they
induce. These decays give rise to novel Higgs decay signals that could
be searched for at colliders. The most interesting searches are final
states containing missing energy and isolated charged leptons such as
$\ell^\pm\ell^\mp, \ell^\pm\ell^\pm, 3\ell$, and $4\ell$.  In addition
to these standard searches, the neutralino decay length can be long
enough to leave a displaced vertex in the detector
\cite{Porod:2000hv,deCampos:2005ri,deCampos:2007bn}. These searches
may facilitate the discovery of the Higgs boson at the LHC and at the
same time provide direct evidence of R-parity violation.

The rest of this paper is organized as follows: In
section~\ref{sec:r-parity-viol} we discuss briefly the model
considered as well as the constraints on the R-parity violating
couplings. Section~\ref{sec:neutralino-decay} is devoted to neutralino
decays and presents our main results.  The decay length is studied and
several scenarios, in which different neutralino final states are
present, are analyzed. Finally in section~\ref{sec:conclusions} we
present our conclusions.

\section{The R-parity violating model}
\label{sec:r-parity-viol}
The most general supersymmetric version of the Standard Model has a
renormalizable superpotential given by
\begin{eqnarray}
  W&=&W_{\mathrm{ MSSM}}+ \varepsilon_{ab} \left[
  \frac{1}{2}\lambda_{ijk}\widehat{L}_{i}^{a}\widehat{L}_{j}^{b}
  \widehat{E}_{k} + \lambda'_{ijk}\widehat{L}_{i}^{a}\widehat{Q}_{j}^{b}
  \widehat{D}_{k} + \epsilon_{i}\widehat{L}_{i}^{a}\widehat{H}_{u}^{b}
  \right]\nonumber\\
  \label{eq:lvsp}
 & &+\varepsilon_{\alpha\beta\sigma}\frac{1}{2}\lambda''_{ijk}
  \widehat{U}_{i}^\alpha\widehat{D}_{j}^\beta
  \widehat{D}_{k}^\sigma~,
\label{eq:superpot}
\end{eqnarray}
where $W_{\mathrm{ MSSM}}$ is the lepton and baryon number conserving
superpotential. In eq.~(\ref{eq:lvsp}), $i,j,k$ run over the fermion
generations, $a,b$ are $SU(2)$ indices, and $\alpha,\beta,\sigma$ are
color indices. The bilinear couplings $\epsilon_i$, $i=1,2,3$, have
mass dimension one and break lepton number. $\lambda$ and $\lambda'$
are dimensionless trilinear couplings that also break lepton
number. There are 27 independent $\lambda'_{ijk}$ but only 9
$\lambda_{ijk}$ --they are antisymmetric in $i,j$. Baryon number is
broken by the dimensionless couplings $\lambda''_{ijk}$, $9$ of which
are independent. In all, therefore, the R-parity violating
superpotential contains $48$ additional parameters. In addition there
are the corresponding soft SUSY breaking terms.

\begin{table}[t]
\begin{center}
\begin{tabular}{cc}
Coupling          & Upper bound\\\hline
$\lambda''_{112}$  & $10^{-7}$\\
$\lambda''_{113}$  & $10^{-5}$\\
$\lambda'_{111}$   & $10^{-4}$\\
$\lambda'_{i33}$   & $10^{-4}$\\
$\lambda_{i33}$    & $10^{-3}$\\
\hline
\end{tabular}
\end{center}
\caption{Strongest bounds on the R-parity violating couplings.}
\label{bounds}
\end{table}

To prevent proton decay, in the MSSM all the operators in
(\ref{eq:lvsp}) are forbidden by assuming the discrete symmetry known
as R-parity. From a phenomenological point of view, however, the
stability of the proton only requires that baryon and lepton number
violating operators not be simultaneously allowed. And R-parity is not
the only discrete symmetry able to ensure that.  Baryon-triality and
lepton-parity~\cite{Ibanez:1991hv}, for instance, are two well
motivated symmetries that allow for \emph{either} lepton or baryon
number violating couplings, breaking R-parity but keeping the proton
stable.

Once R-parity is broken, the stability of the LSP is no longer
guaranteed.  The R-parity breaking terms in eq.~(\ref{eq:lvsp}) in
fact induce the decay of the neutralino -mediated by a gauge boson or
a scalar- to three Standard Model fermions. Such a decay constitutes
the main difference between the MSSM and the R-parity violating model.

Low energy data put a strong bound on some of the R-parity violating
couplings in (\ref{eq:lvsp}). For the bilinear couplings $\epsilon_i$,
the most stringent constraint comes from neutrino physics. The
atmospheric mass scale can be generated at tree level if $\sum_i
(\epsilon_i v_d+\mu v_i)^2/ \mbox{Det}(m_{\chi}) \sim m_\nu/M_2$ where
$\mbox{Det}(m_{\chi})$ is the determinant of the MSSM neutralino mass
matrix, $M_2$ the $SU(2)$ gaugino mass, $v_i$ and $v_d$ are the
sneutrinos and $H_d$ vacuum expectations values. The solar mass scale,
on the other hand, is induced only at the loop level and constrains
the ratio $|\epsilon_i/\mu|\lsim 10^{-3}$.  For the trilinear
couplings the strongest bounds come from double nucleon
decay~\cite{Goity:1994dq}, neutron
oscillations~\cite{Goity:1994dq,Zwirner:1984is}, neutrino
masses~\cite{Godbole:1992fb,Davidson:2000uc}, and neutrinoless double
beta decay~\cite{Hirsch:1995zi}.  A complete and detailed list of the
different constraints on the trilinear couplings can be found
in~\cite{Allanach:1999ic,Barbier:2004ez}. The most important ones are
summarized, for further use, in Table \ref{bounds}.  Note here, that
in principle one can rotate the ''four vector'' $(\hat H_d, \hat L_i)$
such that the bilinear terms are absent in the superpotential
eq.~(\ref{eq:superpot}) on the expense of changing the values of
$\lambda_{ijk}$ and $\lambda_{ijk}'$ (see
e.g.~\cite{AristizabalSierra:2004cy}).  The bounds in this table have
to be understood in this particular basis.

\section{Neutralino decays}
\label{sec:neutralino-decay}
\begin{figure}[t]
\begin{center}
 \includegraphics[scale=0.35,angle=0]{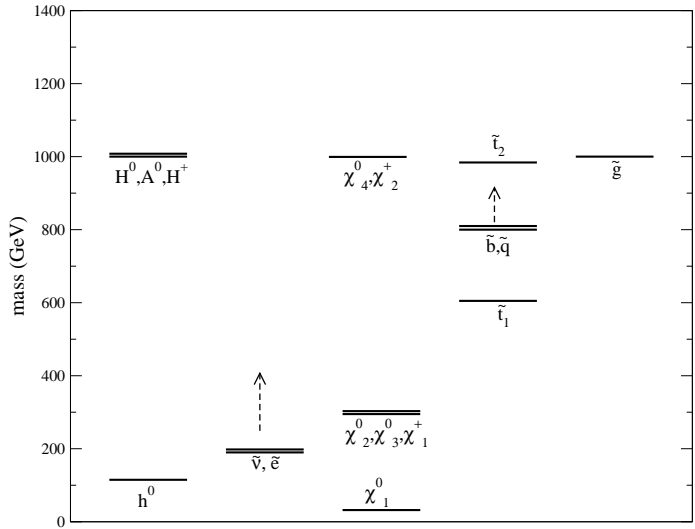}
\caption{The typical supersymmetric spectrum we consider.}
\end{center}
\label{spec}
\end{figure}
 
The branching ratio $\hnn$ depends mainly on $\mu$, $\tan\beta$, and
$m_\chi$ \cite{Yaguna:2007vm}, whereas the subsequent neutralino decay
is determined by $m_\chi$, the $R_p$ violating couplings and gauge
bosons or scalar masses. To study the decay of neutralinos originating
in Higgs decays, we will consider a generic class of supersymmetric
models featuring a non-negligible BR($\hnn$) where all the parameters
but the slepton and the squark masses are kept fixed. We take
\begin{align}
&M_1=35~\gev, \mu=300~\gev, \tan\beta=5\,,\nonumber\\
&M_2=M_3=M_A=1~\tev\,,\nonumber\\
\label{eq:spec}
&A_t=-1.7~\tev\,,\\
&m_{\tilde q}>800~\gev\,,\nonumber\\
&m_{\tilde l}>200~\gev.\nonumber
\end{align}
The supersymmetric spectra thus obtained, illustrated in figure
\ref{spec}, satisfy the constraints from accelerator searches
\cite{Yao:2006px}, the Higgs mass \cite{higgs}, $(g-2)_\mu$
\cite{Yao:2006px,muon}, and $b\to s\gamma$ \cite{bsgamma}. To compute
the spectrum and to evaluate the Higgs mass and other observables we
use the \texttt{FeynHiggs} program \cite{feynhiggs}. According to it,
lighter squarks -with all other parameters fixed- are not compatible
with the LEP bound on the Higgs mass.

Notice that for simplicity we assumed common soft-breaking masses for
sleptons and squarks \footnote{The observed splitting between the two
  stop mass eigenstates is due to the non-zero value of the parameter
  $A_t$} as well as a typical value of $1~\tev$ for $M_2,M_3$ and
$M_A$. Since $M_A\gg M_Z$ we are in fact working in the decoupling
limit, where the interactions of the lightest Higgs boson become
SM-like.

\begin{figure}[t]
\begin{center}
 \includegraphics[width=9cm,height=7cm,angle=0]{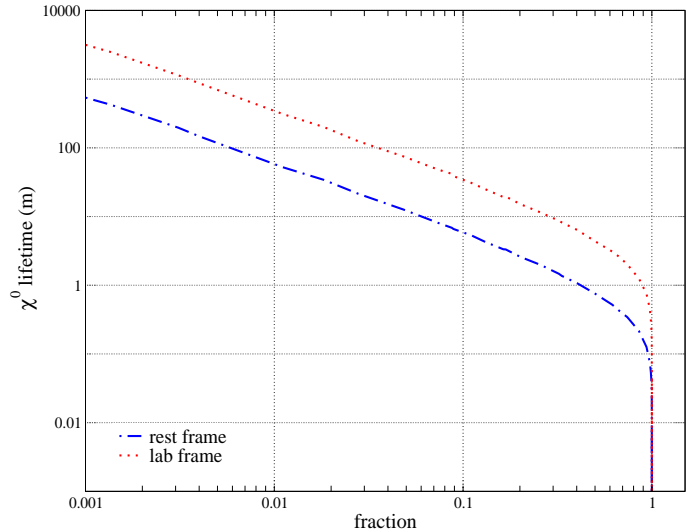}
\end{center}
\caption{The relation between the fraction of neutralinos that decay within the 
detector and  the neutralino decay length considering only
neutralinos from Higgs decays.}
\label{fraction}
\end{figure}

Regarding $\tan\beta$, $\mu$, and $M_1$, they were chosen so as to be
in a region where BR($\hnn$) is non-negligible, and their values are
rather typical within that region. Specifically, we get
BR($\hnn$)$=21\%$, being $b\bar b$ the dominant decay mode with
BR($h\to b\bar b$)$=59\%$. As $M_1\ll M_2,\mu$, the lightest
neutralino is dominantly a bino, but it has a small higgsino component
that generates the non-zero $h\chi\chi$ coupling.

If the neutralino decays outside the detector no new collider signals
due to R-parity violating couplings are expected, as we would
essentially be back to the MSSM case, where the decay $\hnn$ is
invisible. We must, therefore, ensure that a significant number of
neutralinos decays inside the detector.  At the LHC, the Higgs is
mainly produced through gluon fusion and its production cross section
is huge --about $45\,\mathrm{pb}$ for our model. So, the neutralino
decay length could be large and still yield a significant number of
neutralino decays within the detector. This fact is illustrated in
figure \ref{fraction}, where we quantify the fraction of neutralinos
that decay inside a typical detector as a function of the decay
length. This figure was generated with the {\tt PYTHIA} program,
version 6.414 \cite{Sjostrand:2006za}, by taking into account only the
production of the lightest CP-even Higgs. The MSSM parameters were
varied according to eq.~(\ref{eq:spec}) whereas a specific R-parity
violating coupling was varied in the range [$10^{-4}$,$10^{-1}$]. The
points were selected by imposing the condition that the neutralino
decays take place inside a cylinder of 3 m in the $z$ direction and
0.9 m of radius, e.g.~well inside the inner ATLAS or the inner CMS
detector.  From the figure we see, for instance, that if the proper
neutralino lifetime is $10~\mathrm{m}$, about $7\%$ of the decays will
occur within the detector.  Notice therefore that the predicted number
of events will depend on the value of the R-parity violating
couplings.

In what follows we will discuss the non-standard
decays~\cite{Chang:2008cw} of the Higgs boson that are induced by the
R-parity violating couplings.

\subsection{Decays induced by bilinear terms}
\label{sec:bilinear}
Bilinear broken R-parity models are theoretically well motivated
scenarios. They provide a simple framework that accounts for the
observed values of neutrino parameters and that, in contrast with the
seesaw mechanism, can be tested, through the decay properties of the
LSP, in accelerator experiments. These models contain six lepton
number violating parameters \cite{Mira:2000gg,Masiero:1990uj}:
$\epsilon_i$, and their corresponding soft-breaking terms. These
parameters are not entirely free, they are constrained by neutrino
oscillation data.  At present, the experimental data on neutrino
oscillations indicates
that~\cite{Maltoni:2004ei,GonzalezGarcia:2007ib}:
\begin{align}
\tan^2\theta_{12}&=0.47\pm0.05,\qquad  \tan^2\theta_{23}=0.83^{+0.35}_{-0.17},
\qquad  \sin^2\theta_{13}<0.019\nonumber\\
\Delta m_{21}^2&=7.67^{+0.22}_{-0.21}\times10^{-5}\,\text{eV}^2,\qquad
\Delta m_{31}^2=2.46\pm0.15\times10^{-3}\,\text{eV}^2.
\end{align}
In our analysis, we will demand compatibility at the $1\sigma$ level
between these data and the six bilinear parameters.

Neutralino decays in bilinear broken R-parity models are due to the
mixing between neutralinos and neutrinos. The bilinear soft breaking
terms, indeed, induce non-zero vevs for the sneutrinos that give rise
to a mixing between leptons and gauginos and between sleptons and
higgses. Such mixing allows the neutralino to decay -via a $Z^0$,
$W^\pm$, sfermion or Higgs exchange- into the final states
$\nu_i\nu_j\nu_k$, $\nu_i q\bar{q}$, $\nu_i l_j^+ l_k^-$ or $l^\pm
qq'$.

Apart from these contributions induced by the mixing, there are
additional ones due to the \emph{effective} trilinear couplings 
\cite{AristizabalSierra:2004cy}
$\lambda_{233}=h_{\tau}\epsilon_{2}/\mu$ and 
$\lambda'_{333}=h_b \epsilon_3/\mu$.  These new
contributions, mediated by slepton and squarks,  give rise
to neutralino decays into the final states $\nu\tau\tau$, $\nu\mu\tau$, 
and $\nu b\bar b$. 

\begin{figure}[t]
  \centering
  \includegraphics[scale=0.7]{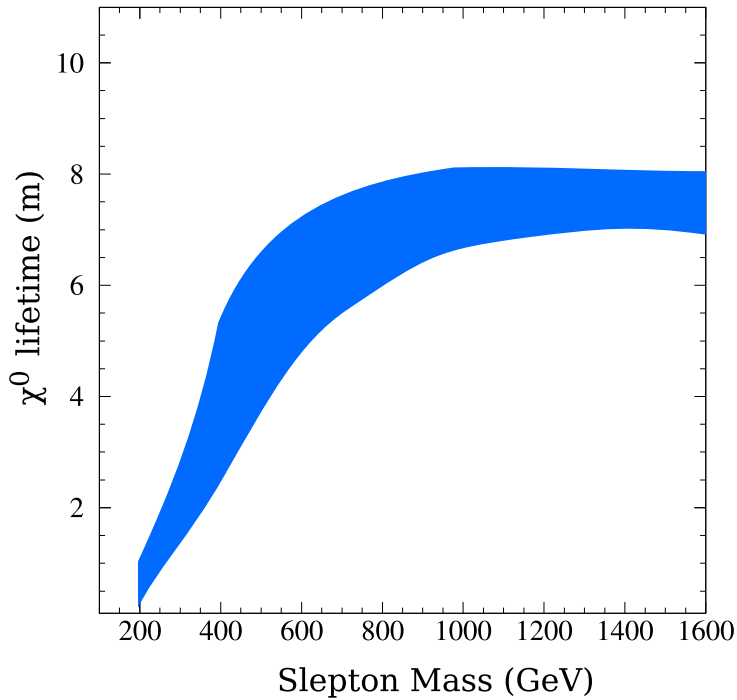}
  \caption{The neutralino lifetime as a function of the slepton mass 
    for different values of the bilinear R-parity violating couplings 
    consistent with neutrino oscillation data at 1 $\sigma$ level.}
  \label{fig:bili_l}
\end{figure}

Figure \ref{fig:bili_l} shows the neutralino lifetime as a function of
the slepton mass for models with bilinear R-parity violation. The
supersymmetric spectrum was chosen according to equation
(\ref{eq:spec}) and the figure was generated with a private version of
\texttt{SPheno}\footnote{This version can be obtained from
  W.P.}~\cite{Porod:2003um} that includes bilinear R-parity
violation. For any given value of the slepton mass, there is a spread
in the neutralino lifetime that is due to the uncertainty on neutrino
parameters.  From the figure we see that the neutralino decay length
has an upper bound of roughly $8$ meters and that it is always larger
than about $50 \mathrm{cm}$. That means that, according to figure
\ref{fraction}, between $7\%$ and $40\%$ of neutralino decays will
occur within the detector.

\begin{figure}[t]
  \centering
  \includegraphics[scale=0.6]{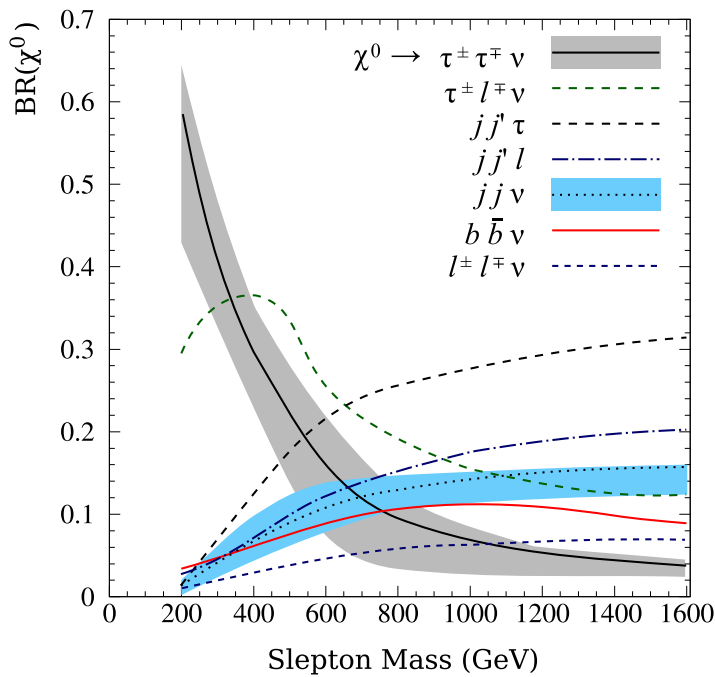}
  \caption{Neutralino decay branching ratios as a function of the 
    slepton mass. The bilinear R-parity violating couplings are 
    consistent with neutrino oscillation data at 1 $\sigma$ level.}
  \label{fig:br}
\end{figure}

Two different regions can be easily distinguished from figure
\ref{fig:bili_l}. For low slepton masses, $m_{\tilde\ell}\lesssim 800
\gev$, the neutralino lifetime increases with the slepton
mass. Neutralino decays in this region are thus mediated by sleptons
and induced by the effective trilinear couplings. For larger slepton
masses, instead, the neutralino lifetime becomes essentially
independent of $m_{\tilde\ell}$.  In this region neutralino decays are
mediated by gauge bosons and induced directly by the mixing.  This
picture is confirmed by figure \ref{fig:br}, where we show the
corresponding neutralino branching ratios as a function of the slepton
mass.  As expected, the dominant decay modes at low slepton masses are
$\tau^\pm\tau^\mp\nu$ and $\tau^\pm l^\mp\nu$ ($l=e,\mu)$ whereas for
large slepton masses several final states have sizeable branching
ratios.

Models with bilinear R-parity violation have thus two remarkable
features.  On the one hand, the neutralino lifetime can be predicted
with certain confidence. It lies between $50$ cm and about $8$ meters;
so it is rather large but it has a known upper bound. On the other
hand, the neutralino decay products are rather uncertain and strongly
depend on the sfermion masses. However, certain rations of branching
ratios are predicted in terms of neutrino mixing angles
\cite{Porod:2000hv}. In general, several final states with sizable
branching ratios are expected.

\subsection{Decays induced by trilinear terms}
The R-parity violating couplings $\lambda,\lambda', \lambda''$ induce
3-body neutralino decays into Standard Model particles.  To study such
decays, we assume that, in turn, only one of the couplings, say
$\lambda_{122}$, dominates and all others are negligible. First we
consider the baryon number violating couplings $\lambda''_{ijk}$, and
then the lepton number violating $\lambda'_{ijk}$ and
$\lambda_{ijk}$. For each case we compute the neutralino lifetime,
$\tau_\chi$, as a function of the couplings and the sfermion
masses. With that information, we determine the range of R-parity
violating parameters that lead to neutralino decays inside the
detector, find the new Higgs decay signals they induce, and briefly
analyze the possibility of observing them at colliders.

For simplicity, we work in the approximation where all final state
fermions are massless\footnote{We have checked that even taking $m_b$
  properly into account changes the total width only slightly.}.  This
approximation breaks down only if there is a top quark in the final
state. Such a decay, however, is not kinematically allowed. The
neutralino decay width, then, has the generic form
\begin{equation}
\label{eq:decay-length}
  \Gamma_\chi=\frac{c_fg^2\{\lambda,\lambda',\lambda''\}^2 
    m_\chi^5}{1536\pi^3} f(m_{\tilde f}),
\end{equation}
where $c_f$ is the color factor, $\{~\}$ denotes one of the couplings,
and $f(m_{\tilde f})$ is a function of dimension $m^{-4}$ that depends
on the masses of the intermediate sfermions~\cite{Dreiner:1994tj}.

\subsubsection{Decays induced by  $\lambda''$}
\label{sec:lambdadprime}
The trilinear coupling $\lambda_{ijk}''$ may induce the decay of
neutralinos into 3-quark final states, such as $\bar u\bar d\bar s$
and $c s b$, leading to a Higgs boson that decays into a six-quark
final state. Such decays were previously considered in
\cite{Carpenter:2007zz}, albeit in a different scenario. Indeed, it
was assumed in \cite{Carpenter:2007zz} that the Higgs boson had a mass
around $100~\gev$ and had been missed by LEP searches because of its
dominant decay into six quarks. Such a situation, however, is only
possible in a very restricted portion of the viable parameter
space. We, instead, consider a more generic framework where the Higgs
is compatible with the usual LEP bound and has a non-dominant
BR($\hnn$).

\begin{figure}[t]
  \begin{center}
    \includegraphics[scale=0.4,angle=0]{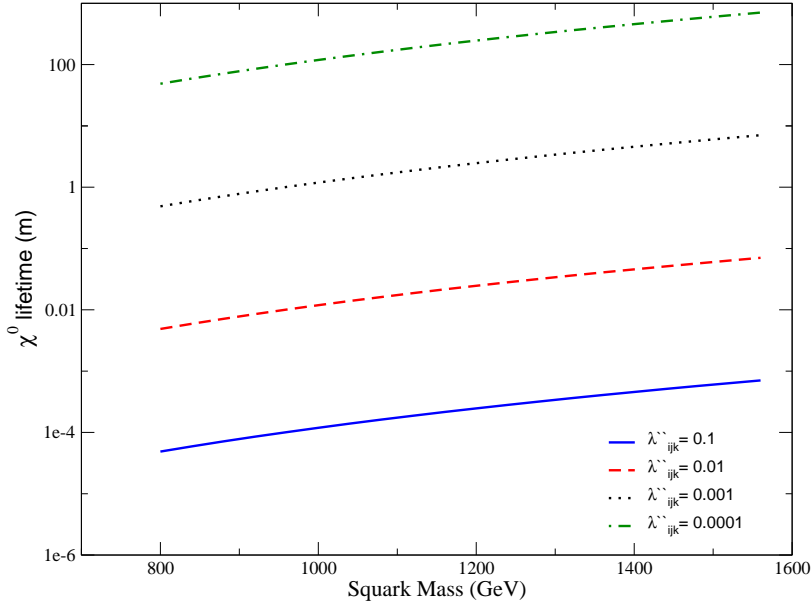}
  \end{center}
  \caption{The neutralino lifetime as a function of the squark mass
    for different values of the R-parity violating couplings
    $\lambda''$.}
\label{ldp}
\end{figure}

Three different diagrams, respectively mediated by $\tilde u_i$,
$\tilde d_j$, and $\tilde d_k$, contribute to the neutralino decay
induced by a given $\lambda''_{ijk}$.  The possible final states are
$u_i d_j d_k $ ($j\ne k$) and its conjugate $\bar u_i\bar d_j\bar
d_k$. Hence, the total decay width is simply given by
$\Gamma_\chi=2\Gamma(\chi\to u_id_jd_k)$.  Besides $\lambda''$,
$\Gamma_\chi$ will only depend on the squark masses. Figure \ref{ldp}
shows the neutralino lifetime as a function of the squark mass for
$\lambda''=10^{-1},10^{-2},10^{-3},10^{-4}$. The resulting neutralino
lifetime varies, for $m_{\tilde q}<1.6~\tev$, approximately between
$100~\mu\mathrm{m}$ and $1~\mathrm{km}$.
 
Notice that the decays induced by the couplings $\lambda''_{3jk}$ are
kinematically forbidden, as they give rise to final states containing
a top quark. Moreover, due to the strong constraint that exist on
$\lambda''_{112}$ and $\lambda''_{113}$, see Table~\ref{bounds},
neutralino decays into $ uds$ and $udb$ are very suppressed and take
place outside the detector.

The unique signal from Higgs decays in this case is then a six-quark
final state. But it is not known whether such signal could be observed
over the QCD background. An interesting possibility, put forward in
\cite{Kaplan:2007ap}, is the search -at LHCb- for the displaced
vertices associated with the neutralino decay. As observed in figure
\ref{ldp}, the neutralino decay length may be larger than
$100\mu\mathrm{m}$, leaving a displaced vertex; and the LHCb detector,
in contrast to ATLAS and CMS, is well suited to study such events.

\subsubsection{Decays induced by  $\lambda'$}
\label{sec:lambdaprime}
Neutralino decays induced by the trilinear coupling $\lambda'$ contain
leptons in the final state and are, therefore, easier to observe.  An
analysis related to ours was presented several years ago in
\cite{Datta:2000ja}. They considered the special case of Higgs
production through vector boson fusion and studied the Higgs signals
induced by only certain trilinear couplings $\lambda$ and $\lambda'$
assuming gaugino mass unification.

The coupling $\lambda_{ijk}'$ gives rise to two different final states
(plus their conjugates):
\begin{align}
\chi&\to e_i u_j\bar d_k\,, \label{eq:lpc}\\
\chi&\to\nu_i d_j\bar d_k.\label{eq:lpn}
\end{align}
And each of them receives contributions from three different
diagrams. The decay (\ref{eq:lpc}) may have $\tilde e_i$, $\tilde
u_j$, and $\tilde d_k$ as intermediate particles while $\tilde \nu_i$,
$\tilde d_j$ and $\tilde d_k$ mediate the process (\ref{eq:lpn}). The
total neutralino decay width is then given by
\begin{equation}
  \label{eq:dpdecaywidth}
  \Gamma_\chi=2(\Gamma(\chi\to e_iu_j\bar d_k)+ 
  \Gamma(\chi\to \nu_i d_j \bar d_k)).
\end{equation}

\begin{figure}[t]
  \begin{center}
    \includegraphics[scale=0.4,angle=0]{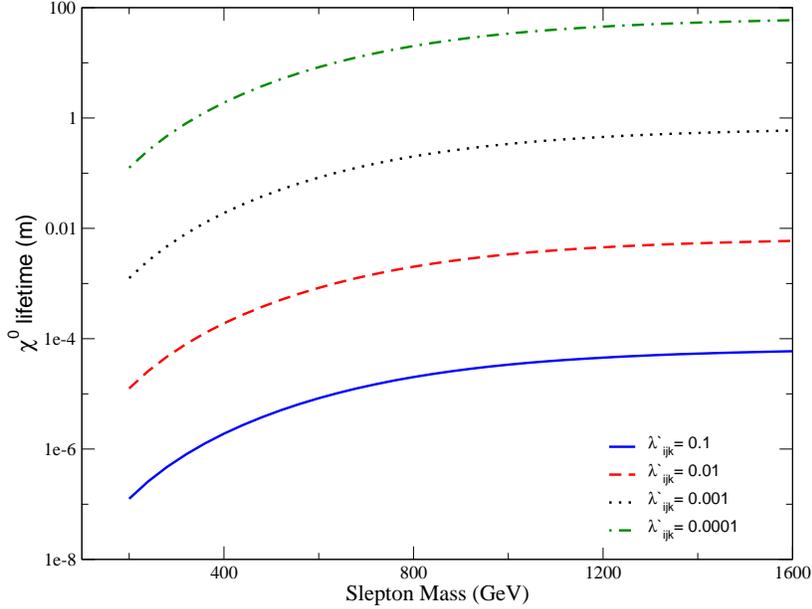}
  \end{center}
  \caption{The neutralino lifetime as a function of the slepton mass
    for different values of the R-parity violating couplings
    $\lambda'$. The common squark mass was set to $800~\gev$.}
\label{lp}
\end{figure}

In this case the neutralino decay width depends on $\lambda'$, the
squark masses, and the slepton masses. We show, in figure \ref{lp},
the neutralino decay length as a function of the slepton mass for
$\lambda'=10^{-1},10^{-2},10^{-3},10^{-4}$ and $m_{\tilde
  q}=800~\gev$.  Notice that for large slepton masses, the diagrams
mediated by squarks tend to dominate over those mediated by sleptons
and consequently the curve becomes rather flat. From the figure we see
that the neutralino lifetime varies, for $m_{\tilde \ell}< 1.6~\tev$,
between $0.1\mu\mathrm{m}$ and $100\mathrm{m}$. Given that the final
state $e_it\bar d_k$ is not kinematically allowed, the couplings
$\lambda'_{i3k}$ induce neutralino decays only into $\nu_i b\bar
d_k$. For those couplings, therefore, the neutralino lifetime is
actually a factor of two larger than shown in figure \ref{lp}.

The possible signatures of the decay of the Higgs boson due to the
coupling $\lambda'$ are:
\begin{enumerate}
\item \emph{Zero lepton, jets, and missing energy.}
\item \emph{One lepton,  jets, and missing energy.}
\item \emph{Opposite sign lepton pair and  jets.}
\item \emph{Same sign lepton pair and jets.}
\end{enumerate}

Since standard searches for supersymmetry at colliders usually rely on
missing energy signals, events with no missing energy, such as
$2\ell+\,$jets, might simply be rejected at the trigger level and
never be available to study. They will, however, give rise to a
displaced vertex provided that $\lambda'\lsim 0.01$. The decays
induced by the couplings $\lambda'_{i3k}$ always give rise to jets
plus missing energy signals. Even after the degrading in missing
energy compared with the case of stable neutralino, the signal with
jets and neutrinos, which has at least a $50\%$ branching, has good
potential to be discovered at the LHC \cite{deCampos:2007bn}.
A generic expectation of this scenario is that the Higgs should be
discovered at LHCb \cite{Kaplan:2007ap}.
 
\subsubsection{Decays induced by  $\lambda$}
\label{sec:lambda}
\begin{figure}[t]
  \begin{center}
    \includegraphics[scale=0.4,angle=0]{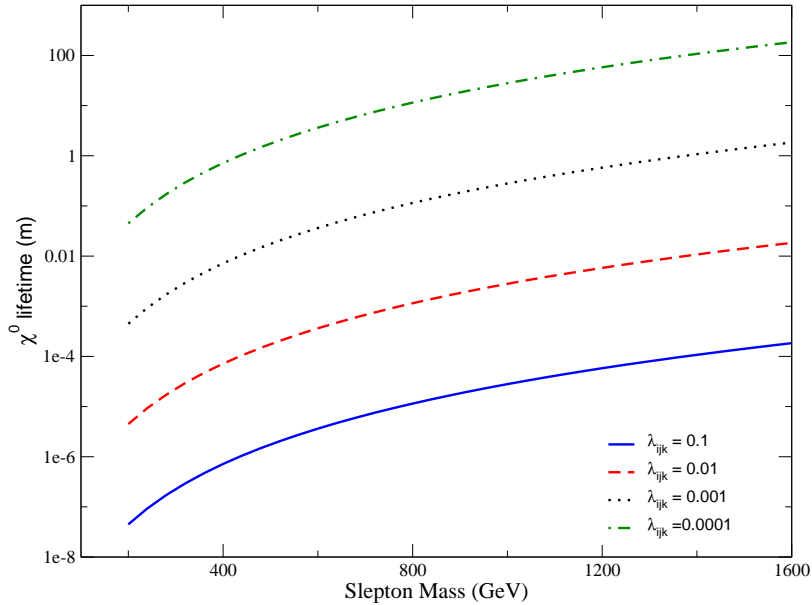}
  \end{center}
  \caption{The neutralino lifetime as a function of the slepton mass for different values of the R-parity 
    violating couplings $\lambda$.}
\label{lam}
\end{figure}

The couplings $\lambda_{ijk}$ induce neutralino decays into final
states containing two charged leptons and one neutrino. A given
$\lambda_{ijk}$ might lead to two different final states (plus their
conjugates):
\begin{align}
\chi&\to e_i\nu_j \bar e_k\\
\chi&\to\nu_i e_j \bar e_k.
\end{align}
Hence, a neutrino is always present in the final state. As before,
three different diagrams contribute to each final state and the
resulting decay width simply scales as $1/m_{\tilde \ell}^4$.  Figure
\ref{lam} shows the neutralino decay length as a function of the
common slepton mass for different values of the coupling $\lambda$.

The two signatures with missing energy we mentioned in
\ref{sec:lambdaprime} -numerals $1$ and $2$- will also be present in
this case, though the jets come from the hadronic decay of the tau
lepton and not from final state quarks. Additionally, the $\lambda$
couplings also give rise to new signatures with two or more
leptons. They are
\begin{enumerate}
\item \emph{Opposite sign lepton pair, jets, and missing energy.}
\item \emph{Same sign lepton pair, jets, and missing energy.}
\item \emph{3 leptons + jets + missing energy.}
\item \emph{4 leptons + missing energy.}
\end{enumerate} 
Notice that the decays induced by $\lambda$, in contrast with those
due to $\lambda'$, always lead to missing energy --from the final
state neutrino.  These new signatures with two or more leptons are
particularly significant because thanks to their low backgrounds they
have good chances to be discovered at LHC
\cite{deCampos:2007bn,Baer:1992kd}.

\section{Conclusions}
\label{sec:conclusions}
We studied the decay of the Higgs boson into Standard Model particles
within the context of the MSSM with R-parity violation. The decay
proceeds via $\hnn$ followed by the R-parity violating neutralino
decay into light fermions. We pointed out that neutralino decays
induced by the trilinear R-parity violating couplings may occur inside
the detector and, as a result, give rise to novel Higgs decay
signatures that may facilitate the discovery of the Higgs boson at the
LHC.  Particularly appealing -because of their low backgrounds- are
the decays into final states containing three or four leptons and
missing energy. Such decays are caused by the lepton-number violating
couplings $\lambda_{ijk}$ and could be easily observed at the LHC.
Another promising possibility is the observation of displaced vertices
associated with the neutralino decay. Thus, the discovery of the Higgs
boson at the LHC might also provide direct evidence of R-parity
violation.

\section{Acknowledgments}
Work partially supported by Colciencias in Colombia under contract
1115-333-18740. D.A.S. is supported by an INFN posdoctoral
fellowship. C. Y. is supported by the \emph{Juan de la Cierva} program
of the Ministerio de Educacion y Ciencia de Espana. W.P. is partially
supported by the German Ministry of Education and Research (BMBF)
under contract 05HT6WWA.  D.A.S. and C.Y. thank the particle group at
Universidad de Antioquia for their hospitality during the completion
of this work.

\end{document}